\documentclass{article}
\usepackage{amsmath,amsthm}
\usepackage{amssymb,latexsym}
\usepackage[mathscr]{eucal}
\usepackage{setspace}
\usepackage{graphics}
\usepackage{array}

\newcommand{\lp}{\left(}
\newcommand{\rp}{\right)}

\newcommand{\be}{\begin{equation}}
\newcommand{\ee}{\end{equation}}

\begin{document}

\title{\textbf{Singularities in Fully Developed Turbulence}}         
\author{Bhimsen K. Shivamoggi\\
University of Central Florida\\
Orlando, FL 32816-1364, USA
}        
\date{}          
\maketitle

\noindent{\Large{\bf Abstract}}

\vspace{.10in}

\large Phenomenological arguments are used to explore finite-time singularity (FTS) development in different physical fully-developed turbulence (FDT) situations. The role played by the cascade physics underlying the FTS development is investigated. Such diverse aspects as the effects of spatial intermittency and fluid compressibility in three-dimensional (3D) FDT and the role of the divorticity amplification mechanism in two-dimensional (2D) FDT and quasi-2D quasi-geostrophic FDT and the advection-diffusion mechanism in magnetohydrodynamic turbulence are considered to provide physical insights into the FTS development in variant cascade physics situations. The quasi-geostrophic FDT results connect with the 2D FDT results in the barotropic limit while they connect with 3D FDT results in the baroclinic limit (on doing the necessary interchange of vorticity in the 3D case with divorticity in the quasi-2D case); hence they seem to provide a kind of bridge between 2D FDT and 3D FDT results.

\pagebreak

\noindent\Large{\bf 1. Introduction}\\

\large The existence of strongly localized features like vorticity sheets in the small-scale structure of three-dimensional (3D) fully-developed turbulence (FDT) suggests the development of singularities in the flow variables\footnote{The mechanism of vorticity stretching and amplification typically involves the generation of small-scale structures (example: the Burgers vortex).} (Gibbon \cite{Gib}). On the mathematical side, the 3D Euler equations may be viewed as a quadratic evolution equation for vorticity $\boldsymbol{\omega}$ which, in principle, allows some solutions to evolve from smooth initial data to a singularity in {\it finite} time (Constantin \cite{Con}), so this possibility is of interest to mathematicians in the context of global existence of solutions. It is generally believed that an understanding of this process\footnote{One of the issues in this regard is the connection, if any, between the energy transfer to small scales and {\it dissipative anomaly} associated with the time-asymmetry property in FDT and the FTS.} is crucial to the development of a viable theory of turbulence (Constantin \cite{Con}). A rigorous result due to Beale et al. \cite{Bea} requires the magnitude of the vorticity $\boldsymbol{\omega}$ to become infinite to allow the occurrence of a {\it finite-time} singularity (FTS)\footnote{$\boldsymbol{\omega}$ amplifies when it roughly aligns with the linear eigenspace associated with the positive intermediate eigenvalue of the local flow-strain matrix (Majda \cite{Maj2}) while the latter is found to be associated with the time-asymmetry property in FDT (Jucha et al. \cite{Jucha}).}. More specifically, this result states that a singularity in a flow variable develops as $t \to t^*$, only if
\be\tag{1a}
\int_0^t \Vert \omega \Vert_{\infty} (\tau) \ d\tau \to \infty \ , \ \text{as } t \to t^* \ .
\ee
This result implies
\be\tag{1b}
\Vert \omega \Vert_{\infty}(t) \sim \frac{1}{(t^* -t)^{\beta}} \ , \ \beta \geq 1 \ .
\ee
Here, $\Vert \omega \Vert_{\infty} \equiv \max\limits_{\boldsymbol{x} \in \mathbb{R}^3} \Vert \omega(\boldsymbol{x}) \Vert$.

 Numerical investigations of 3D FDT (see Gibbon \cite{Gib} for a detailed summary) have suggested but failed to provide a conclusive evidence that ideal-flow solutions, starting from regular initial conditions, will spontaneously develop a singularity in {\it finite} time (for example, Brachet et al. \cite{Bra}). Mathematically, it is not yet resolved if the Euler equations are guaranteed to possess solutions that remain smooth, for all $t$, if the initial data are smooth. The existing proofs establish either existence of smooth solutions, which is local in $t$, or weak solutions, for all $t$, whose smoothness is not guaranteed. The outstanding open issue therefore is (Majda and Bertozzi \cite{Maj}), ``Are there smooth solutions with finite energy of the 3D Euler equations that develop singularities in a finite time?''

Phenomenological considerations, which predict development of FTS in FDT, are believed to over-estimate the nonlinear effects (Frisch \cite{Fri}) because nonlinearity depletion mechanisms via local alignment of vorticity and the consequent coherent structure generation seem to be operational (Constantin \cite{Con}, Frisch \cite{Fri}) and enhanced coherence of the vorticity field makes the conditions for the singularity development more stringent. Nonetheless, here we propose to use phenomenological arguments to explore the FTS development in different physical FDT situations because, from a qualitative point of view, such considerations do seem to be able to provide useful physical insights into this process in variant cascade physics situations.

\vspace{.3in}
\pagebreak

\noindent\Large{\bf 2. 3D Incompressible FDT}\\

\large Vortex stretching and amplification is believed to constitute the physical mechanism underlying the energy cascade in 3D FDT. So, the vorticity $\boldsymbol{\omega} \equiv \nabla \times {\bf v}$ is the appropriate physical variable to characterize the dynamics in question. An estimate for the energy dissipation rate $\epsilon$ in 3D FDT is given by,
\be\tag{2}
\varepsilon \sim \nu \frac{v^2}{\eta^2}.
\ee
$v$ being the velocity increment over the {\it Kolmogorov microscale} $\eta$ and $\nu$ being the kinematic viscosity. On using the Kolmogorov scaling, 
\be\tag{3}
v \sim \varepsilon^{1/3} \eta^{1/3}
\ee
(2) gives the well-known result, 
\be\tag{4}
\eta \sim \frac{\nu^{3/4}}{\varepsilon^{1/4}}.
\ee

On using (4), the vorticity evolution equation,
\be\tag{5}
\frac{d \boldsymbol{\omega}}{dt} = ( \boldsymbol{\omega} \cdot \nabla) {\bf v} + \nu \nabla^2 \boldsymbol{\omega}
\ee
leads to the vorticity growth rate estimate,
\be\tag{6}
\frac{d \omega}{dt} \sim \nu \frac{\omega}{\eta^2} \sim \frac{\varepsilon^{1/2}}{\nu^{1/2}} \omega.
\ee

Invoking the {\it dissipative anomaly}, in 3D FDT,
\be\tag{7}
\varepsilon \sim \nu \omega^2 \sim const
\ee
(6) leads to 
\be\tag{8}
\frac{d \omega}{dt} \sim \omega^2
\ee
and hence the well-known result (Frisch \cite{Fri}), 
\be\tag{9}
\omega \sim \frac{1}{t + c}
\ee
suggesting a FTS in 3D FDT. Here, $c$ is an arbitrary constant.

\vspace{.3in}

\noindent\Large{\bf 3. Effects of Spatial Intermittency in 3D FDT}\\

\large Spatial intermittency effects associated with the spatial spottiness of the turbulent activity become more pronounced at small scales. So, one may surmise spatial intermittency to have a significant effect on the FTS development. One may then incorporate spatial intermittency effects, following Mandelbrot \cite{Mon}, via the fractal nature of strongly convoluted dissipative structures where the turbulent activity is concentrated. In a first approximation, the dissipative structures may be approximated by a homogeneous fractal\footnote{In a homogeneous fractal model (also called the $\beta$-model) for the dissipative structures, the energy flux is assumed to be transferred to only a fixed function $\beta$ of the eddies downstream in the cascade (Frisch et a. \cite{Fri2}).} with a non-integer Hausdorff dimension $D_0$ (Frisch et al. \cite{Fri2}). 

Assuming the scaling behavior\footnote{The non-differentiability of the velocity field implies that the H\"{o}lder scaling exponent $\alpha$ satisfies the condition $\alpha \leq 1/3$ (as confirmed by (13) below).},
\be\tag{10}
v \sim \eta^\alpha
\ee
we have, 
\be\tag{11}
\varepsilon \sim \eta^{3 \alpha - 1}.
\ee

Using (11), (4) gives (Paladin and Vulpiani \cite{Pal}, Sreenivasan and Meneveau \cite{SM}),
\be\tag{12}
\eta \sim \nu^{\frac{1}{1 + \alpha}}.
\ee

The homogeneous fractal model for the 3D FDT (Frisch et al. \cite{Fri2}) gives\footnote{Since $D_0 \leq 3$, (13) implies that the velocity field singularities are strengthened by spatial intermittency.}
\be\tag{13}
\alpha = \frac{1}{3} \lp D_0 - 2 \rp.
\ee

Using (13), (12) gives
\be\tag{14}
\eta \sim \nu^{\frac{3}{D_0 + 1}}.
\ee

Using (14), (5) leads to the vorticity growth rate estimate,
\be\tag{15}
\frac{d \omega}{dt} \sim \nu \frac{\omega}{\eta^2} \sim \nu^{\frac{D_0 - 5}{D_0 + 1}} \omega.
\ee

Using (7), (15) becomes
\be\tag{16}
\frac{d \omega}{dt} \sim \omega^{\frac{11 - D_0}{D_0 + 1}}
\ee
from which,
\be\tag{17}
\omega \sim \lp t + c \rp^{-1 + \frac{3}{2} \lp \frac{D_0 - 3}{D_0 - 5} \rp}.
\ee

The weakening of the FTS due to spatially intermittency $\lp D_0 < 3 \rp$, indicated by (16), reflects a nonlinearily depletion activity occurring in the latter case via generation of coherent structures\footnote{Mailybaev \cite{Maily2} suggests, by considering a shell model for FDT, that the coherent structure generation may be induced by the FTS.} and the consequent enhanced coherence of the vorticity field, as conjectured by Frisch \cite{Fri}.

\vspace{.3in}

\noindent\Large{\bf 4. Effects of Compressibility}\\

\large Compressibility effects on FDT are of great importance in modern technological as well as astrophysical flows (Shivamoggi \cite{Shi} and other references given thereof). There is an intuitive belief that vortices tend to become more resilient and stretch stronger in a compressible fluid. It is therefore pertinent to explore the effect of fluid compressibility on the FTS development.

For the compressible case, an estimate for the kinetic energy dissipation rate $\hat{\varepsilon}$ is given by
\be\tag{18}
\hat{\varepsilon} \sim \frac{\rho v^3}{\hat{\eta}} \sim \mu \frac{v^2}{\hat{\eta^2}}
\ee
from which, we obtain
\be\tag{19a, b} \left. \begin{array}{cl}
\hat{\eta} \sim & \displaystyle \frac{\mu}{\rho v} \\
v \sim & \lp \displaystyle \frac{\hat{\varepsilon} \hat{\eta}}{\rho} \rp^{1/3}
\end{array} \right\}
\ee
and hence (Shivamoggi \cite{Shi}), the Kolmogorov microscale $\hat{\eta}$ for compressible FDT is given by 
\be\tag{20}
\hat{\eta} \sim \lp \frac{\mu^3}{\rho^2 \hat{\varepsilon}} \rp^{1/4}. 
\ee

The vorticity growth rate estimate,
\be\tag{21}
\rho \frac{d\omega}{dt} \sim \mu \frac{\omega}{\hat{\eta}^2}
\ee
on using (20), becomes
\be\tag{22}
\frac{d \omega}{d t} \sim \frac{\hat{\varepsilon}^{1/2}}{\mu^{1/2}} \omega.
\ee

Noting the dissipative anomaly for the compressible case (Shivamoggi \cite{Shi2}),
\be\tag{23}
\hat{\varepsilon} \sim \mu \omega^2 \sim const
\ee
(22) becomes
\be\tag{24}
\frac{d \omega}{d t} \sim \omega^2
\ee
and hence
\be\tag{25}
\omega \sim \frac{1}{t + c}
\ee
as in the incompressible case. The apparent absence of a compressibility correction to (9) is probably due to the fact that compressibility effects are not length-scale dependent and materialize equally at all length scales unlike viscous effects.

\indent On the other hand, in the ultimate compressibility limit, the dissipative structures become {\it shock waves}, which may be associated with the development of singularities in the flow variables in a compressible FDT (Mailybaev \cite{Maily}). In this limit, we have (Shivamoggi \cite{Shi})\footnote{It may be noted that (26a, b) lead to the following scaling behavior for the kinetic energy per unit mass,
\be\notag
\rho v^2 \sim \hat{\eta}
\ee
which, in turn, leads to the Kadomtsev-Petviashvili \cite{Kadom} spectral law for compressible FDT,
\be\notag
E(k) \sim k^{-2}
\ee
$E$ being the kinetic energy spectral density and $k$ being the wave number.},
\be\tag{26a, b, c}
\left. \begin{array}{rl}
\rho \sim & \hat{\eta} \\
v \sim & \text{const} \\
\omega \sim & \hat{\eta}^{-1}
\end{array} \right\} .
\ee
\indent Using (26a, b, c), (21) leads to
\be\tag{27}
\frac{d\omega}{dt} \sim (\mu \omega^2 ) \ \omega^2
\ee
and using (23), (27) leads to
\be\tag{28}
\frac{d\omega}{dt} \sim \omega^2
\ee
in agreement with (24).

\vspace{.3in}

\noindent\Large{\bf 5. Enstrophy Cascade in 2D FDT}\\

\large As a further example of the effect of the underlying cascade physics on the FTS development, let us consider the enstrophy cascade in 2D FDT. 2D FDT is relevant to large-scale atmospheric and oceanic flows and differs from 3D FDT in that it does not have the vortex stretching mechanism operational in 3D (see Tabeling \cite{Tab} for a recent review)\footnote{The vorticity evolution equation in 2D,
\be\notag
\frac{d \boldsymbol{\omega}}{dt} = {\bf 0}
\ee
implies the absence of the vortex stretching mechanism in 2D and leads to the {\it Lagrange invariant},
\be\notag
\boldsymbol{\omega} = \text{const.}
\ee
This invariant introduces a strong restriction on the dynamics underlying 2D.}.

Noting that divorticity (Kida \cite{Kida}) amplification constitutes the physical mechanism underlying the enstrophy cascade in 2D FDT (Kuznetsov et al. \cite{Kuz}) the divorticity {\bf b},
\be\tag{29}
{\bf b} \equiv \nabla \times \boldsymbol{\omega}
\ee
appears to be the appropriate physical variable to characterize the dynamics in question (Shivamoggi et al. \cite{Shi3}). 

The divorticity evolution equation
\be\tag{30}
\frac{d {\bf b}}{d t} = ( {\bf b} \cdot \nabla ) {\bf v} + \nu \nabla^2 {\bf b} 
\ee
on noting that the {\it Kraichnan microscale} $\zeta$ (which is the 2D FDT counterpart of the Kolmogorov microscale for 3D FDT) is given by (Shivamoggi \cite{Shi4}), 
\be\tag{31}
\zeta \sim \frac{\nu^{1/2}}{\tau^{1/6}} 
\ee
leads to the divorticity growth rate estimate,
\be\tag{32}
\frac{d b}{d t} \sim \nu \frac{b}{\zeta^2} \sim \tau^{1/3} b.
\ee
Here, $\tau$ is the enstrophy dissipation rate. (32) further leads to
\be\tag{33}
b \sim e^{\tau^{1/3} t}
\ee
confirming the well known absence of a FTS in 2D FDT (Rose and Sulem \cite{Ros}) consequent to the global-in-time regular behavior of 2D Navier-Stokes solutions evolving from smooth initial data in the inviscid limit. 

\vspace{.3in}

\noindent\Large{\bf 6. Enstrophy Cascade in Quasi-geostrophic FDT}\\

\large As indicated by the development in Section 5, the Lagrange invariance of vorticity introduces a strong restriction on the dynamics underlying 2D flows. It is of interest to consider the effect of a violation of this property on the FTS development. As an example, consider the enstrophy cascade in quasi-geostrophic\footnote{Quasi-geostrophic dynamics refers to the nonlinear dynamics governed by the first-order departure from the linear geostrophic balance between the Coriolis force and pressure gradient transverse to the rotation axis of a rapidly rotating fluid. Quasi-geostrophic FDT refers to randomly varying flow states of fluids that lie close to a state of geostrophic and hydrostatic balance (Charney \cite{Cha}).} FDT, which is relevant to large-scale oceanic flows with the inclusion of the Coriolis force and free-surface effects. The possibility of a FTS in the quasi-geostrophic model and its variants has been numerically discussed by Hoyer and Sadourny \cite{Hoy}, Constantin et al. \cite{Con2}, Ohkitani and Yamada \cite{Ohk}, Scott and Dritschel \cite{Sco}.

For this case, the Kraichnan microscale $\zeta$ is given by (see Appendix),
\be\tag{34a, b}
\zeta \sim \left\{
\begin{array}{ll}
\displaystyle{\frac{\nu^{1/2}}{\tau^{1/6}}} & , \ \zeta \ll R \\
\displaystyle{\frac{\nu^{3/4}}{R^{1/2} \tau^{1/4}}} & , \ \zeta \gg R \ .
\end{array}
\right.
\ee
Here, $R$ is the Rossby deformation radius $R \equiv \sqrt{gH}/f$, $f$ is the local Coriolis parameter, $H$ is the depth of the ocean (taken to be constant), and $g$ is the acceleration due to gravity. We are using the simplest mathematical model of large-scale, nearly horizontal oceanic motion incorporating the force of gravity and the Coriolis force due to the Earth's rotation, which is the one-layer homogeneous ocean with a uniform depth and a spherical free surface. (34a), which corresponds to the barotropic regime, is the same as the 2D FDT result (31), while (34b), which corresponds to the baroclinic regime, is very similar to the 3D FDT result (4). This implies that the quasi-geostrophic FDT results have the potential to connect, in some sense, with the 2D FDT results in the barotropic limit and with the 3D FDT results in the baroclinic limit. This is confirmed in the following. 

Using (34a, b), equation (30) leads to the divorticity growth rate estimate,
\be\tag{35a, b}
\frac{db}{dt} \sim \left\{
\begin{array}{ll}
\tau^{1/3} b & , \ \zeta \ll R \\
\displaystyle{\frac{R \tau^{1/2}}{\nu^{1/2}}} b & , \ \zeta \gg R \ .
\end{array}
\right.
\ee
\par Invoking the dissipative anomaly in quasi-2D FDT,
\be\tag{36}
\tau \sim \nu b^2 \sim const \ ,
\ee
(35a, b) leads to
\be\tag{37a, b}
\frac{db}{dt} \sim \left\{
\begin{array}{lr}
\tau^{1/3} b & , \ \zeta \ll R \\
R b^2 & , \ \zeta \gg R
\end{array}
\right.
\ee
and hence,
\be\tag{38a, b}
b \sim \left\{
\begin{array}{lr}
e^{\tau^{1/3} t} & , \ \zeta \ll R \\
\displaystyle{\frac{1}{Rt + c}} & , \ \zeta \gg R
\end{array}
\right.
\ee
suggesting, in the baroclinic limit, a FTS of the same type as that for vorticity in 3D FDT, namely, (9). Physically, this is due to the enhanced vortex stretching in this limit produced by the deformed free surface in the quasi-geostrophic dynamics.

Thus, the quasi-geostrophic FDT results connect with the 2D FDT results, as to be expected, in the barotropic limit while they connect with the 3D FDT results in the baroclinic limit (on doing the necessary interchange of vorticity in the 3D case with divorticity in the quasi-2D case). Contingent on the latter qualification, the quasi-geostrophic FDT seems to provide some kind of link between the 2D FDT and the 3D FDT. 

\vspace{.3in}

\noindent\Large{\bf 7. Magnetohydrodynamic Turbulence}\\

\large Magnetohydrodynamic (MHD) FDT is relevant to plasmas in astrophysical systems as well as fusion reactors (see Carbone and Pouquet \cite{Carb} for a recent review). In view of the advection-diffusion mechanism that controls the statistical properties of magnetic field\footnote{Advective stretching of magnetic field lines leads to amplification of magnetic field (Batchelor \cite{Bat}) while magnetic field lines that have been highly stretched typically experience stronger Ohmic dissipation (Kida et al. \cite{Kida2}).} the MHD FDT (Shivamoggi \cite{Shi5}) presents a convenient framework to explore the role of the advection-diffusion mechanism in the FTS development. Here, we adopt the Iroshnikov \cite{Iro} - Kraichnan \cite{Kra} (IK) phenomenology. 

The energy dissipation rate in the IK model for MHD FDT is given by
\be\tag{39}
\varepsilon \sim \frac{v^4}{\tilde{\eta} C_A} \sim \eta_m \frac{v^2}{\tilde{\eta}^2}
\ee
where $\eta_m$ is the magnetic resistivity, $C_A$ is the velocity of Alfv\'{e}n waves in the magnetic field of the large-scale eddies, and $\tilde{\eta}$ is the Kolmogorov microscale for MHD FDT. We obtain from (39),
\be\tag{40a, b}
\left. \begin{array}{rl}
\tilde{\eta} \sim & \displaystyle \frac{\eta_m C_A}{v^2} \\
v \sim & \varepsilon^{1/4} C_A^{1/4} \tilde{\eta}^{1/4}
\end{array} \right\} .
\ee
 Combining (40a, b), we obtain
\be\tag{41}
\tilde{\eta} \sim \lp \frac{\eta^2_mC_A}{\varepsilon} \rp^{1/3}.
\ee

The current density $J$ growth rate estimate,
\be\tag{42}
\frac{d J}{d t} \sim \eta_m \frac{J}{\tilde{\eta}^2}
\ee
on using (41), becomes
\be\tag{43}
\frac{d J}{d t} \sim \frac{\varepsilon^{2/3}}{C_A^{2/3} \eta_m^{1/3}} J.
\ee

Noting the dissipative anomaly for MHD turbulence (Shivamoggi \cite{Shi5}),
\be\tag{44}
\varepsilon \sim \eta_m J^2 \sim const
\ee
(43) becomes
\be\tag{45}
\frac{d J}{d t} \sim \frac{\varepsilon^{1/3}}{C_A^{2/3}} J^{5/3}.
\ee

\noindent (45) leads to 
\be\tag{46}
J \sim \frac{1}{\lp t + c \rp^{3/2}}
\ee
showing a strengthening of the FTS in the IK phenomenology, which is plausible because the MHD flows are known to be more dissipative than their hydrodynamic counterparts (Orszag and Tang \cite{Ors}).

\vspace{.3in}

\noindent\Large{\bf 8. Discussion}\\

\large In this paper phenomenological arguments have been used to provide some physical insights into the FTS development in different physical FDT situations. Particular attention is paid to the role played by the cascade physics underlying the FTS development. Such diverse aspects as the effects of spatial intermittency and fluid compressibility in 3D FDT and the role of the divorticity amplification mechanism in 2D FDT and quasi-2D quasi-geostrophic FDT and the advection-diffusion mechanism in MHD turbulence are considered to gain better physical understanding of the FTS development. The quasi-geostrophic FDT results connect with the 2D FDT results, as to be expected, in the barotropic limit while they connect with the 3D FDT results in the baroclinic limit (on doing the necessary interchange of vorticity in the 3D case with divorticity in the quasi-2D case). In this sense, the quasi-geostrophic FDT appears to provide a kind of bridge between the 2D FDT and the 3D FDT. 

\vspace{.3in}

\noindent\Large{\bf Appendix}\\

\large Noting that the potential enstrophy for quasi-geostrophic flows in the Charney \cite{Cha} model is given by
\be\tag{A.1}
U \sim \frac{\phi^2}{\ell^2} \left( \frac{1}{\ell^2} + \frac{1}{R^2} \right)
\ee
the enstrophy transfer rate is given by
\be \tag{A.2a, b}
\tau \sim \left\{
\begin{array}{lr}
\displaystyle{\frac{\phi^3}{\ell^6}} & , \ \ell \ll R \\
\displaystyle{\frac{\phi^3}{R^2 \ell^4}} & , \ \ell \gg R
\end{array}
\right.
\ee
$\phi$ being the stream function.

Assuming that the enstrophy transfer rate is constant in the inertial range, we obtain from (A.2a, b),
\be\tag{A.3a, b}
\phi \sim \left\{
\begin{array}{ll}
\tau^{1/3} \ell^2 & , \ \ell \ll R \\
\tau^{1/3} R^{2/3} \ell^{4/3} & , \ \ell \gg R .
\end{array}
\right.
\ee
On the other hand, using (A.1), the enstrophy dissipation rate is given by
\be\tag{A.4a, b}
\tau \sim \left\{
\begin{array}{lr}
\nu \displaystyle{\frac{\phi^2}{\zeta^6}} & , \ \zeta \ll R \\
\nu \displaystyle{\frac{\phi^2}{R^2 \zeta^4}} & , \ \zeta \gg R
\end{array}
\right.
\ee
$\zeta$ being the Kraichnan microscale.

Using (A.3a, b), (A.4a, b) becomes
\be\tag{A.5a, b}
\tau \sim \left\{
\begin{array}{lr}
\displaystyle{\frac{\nu \tau^{2/3}}{\zeta^2}} & , \ \zeta \ll R \\
\displaystyle{\frac{\nu \tau^{2/3}}{R^{2/3} \zeta^{4/3}}} & , \ \zeta \gg R
\end{array}
\right.
\ee
from which,
\be\tag{A.6a, b}
\zeta \sim \left\{
\begin{array}{ll}
\displaystyle{\frac{\nu^{1/2}}{\tau^{1/6}}} & , \ \zeta \ll R \\
\displaystyle{\frac{\nu^{3/4}}{R^{1/2} \tau^{1/4}}} & , \ \zeta \gg R \ 
\end{array}
\right.
\ee 
as mentioned in (34a, b).

\vspace{.3in}

\noindent\Large{\bf Acknowledgements}\\

\large Some of this work was carried out during the course of my participation in the Turbulence Workshop at the Kavli Institute for Theoretical Physics, Santa Barbara. I am thankful to Professors Katepalli Sreenivasan, Grisha Falkovich and Eberhard Bodenschatz for the invitation. I am thankful to Professor Jens Juul Rasmussen for the valuable suggestions. This research was supported in part by NSF grant No. PHY05-51164.

\end{document}